\documentclass[prl,superscriptaddress,amsmath,amssymb,twocolumn]{revtex4-2}
\usepackage{graphicx}
\usepackage{tcolorbox}
\usepackage{epsfig}
\usepackage{dcolumn}
\usepackage{bbold}
\usepackage{bm}
\usepackage{rotating}
\usepackage{multirow}
\usepackage{xspace}
\usepackage{hyperref} 
\usepackage[english]{babel}

\def\ve{{\varepsilon}}
\def\w{\omega}
\def\bk{{\bf k}}

\def\br{{\bf r}}
\def\bq{{\bf q}}

\newcommand{\ket}[1]{ | #1 \rangle }

\def\>{\rangle}
\def\<{\langle}
\def\D{\partial}

\def\k{\kappa}

\begin{document}
\title{Watching Polarons Form in Real Time}
\author{Victor Garcia-Herrero}
\author{Christoph Emeis}
\affiliation{Institut f\"ur Theoretische Physik und Astrophysik, Christian-Albrechts-Universit\"at zu Kiel, Kiel, Germany} 
\author{{Zhenbang Dai}}
\affiliation{Oden Institute for Computational Engineering and Sciences, The University of Texas at Austin, Austin, TX 78712}
\author{Jon Lafuente-Bartolome}
\affiliation{Department of Physics, University of the Basque Country UPV/EHU, 48940 Leioa, Basque Country, Spain}

\affiliation{Department of Physics, The University of Texas at Austin, Austin, TX 78712}
\author{Feliciano Giustino}
\affiliation{Oden Institute for Computational Engineering and Sciences, The University of Texas at Austin, Austin, TX 78712}
\affiliation{Department of Physics, The University of Texas at Austin, Austin, TX 78712}
\author{Fabio Caruso}
\affiliation{Institut f\"ur Theoretische Physik und Astrophysik, Christian-Albrechts-Universit\"at zu Kiel, Kiel, Germany} 
\begin{abstract}
Polaron formation in pump–probe experiments is an inherently non-equilibrium
phenomenon, driven by the ultrafast coupled dynamics of electrons and phonons,
and culminating in the emergence of a localized quasiparticle state.  In this
work, we present a first-principles quantum-kinetic theory of polaron formation
that captures the real-time evolution of electronic and lattice degrees of
freedom in presence of electron-phonon coupling. We implement this framework to
investigate the ultrafast polaron formation in the prototypical polar insulator
MgO. This approach allows us to determine the characteristic timescales of
polaron localization and to identify its distinctive dynamical fingerprint.
Our results establish clear and experimentally accessible criteria for
identifying polaron formation in pump–probe experiments.
\end{abstract}

\maketitle

Polarons are composite quasiparticles arising from the mutual localization of a
charge carrier and a lattice distortion and represent one of the paradigmatic
manifestations of electron-phonon physics in materials
\cite{franchini_polarons_2021}.  Polarons leave distinctive identifiable
features across a variety of experimental techniques that probe crystals under
static or quasi equilibrium conditions.  Both small and large polarons exhibit
clear fingerprints in temperature-dependent resistivity measurements
\cite{Zhang_2007, Wright_2021}. They further manifest in phenomena such as the
Stokes shift and broad luminescence arising from self-trapped excitons
\cite{Luo_2018}. More recently, real-space localization of polarons has been
directly observed through scanning tunneling microscopy \cite{Liu2023},  and
atomic force microscopy techniques \cite{Redondo_2024}.

\begin{figure}[t]
\centering
\includegraphics[width=0.85\linewidth]{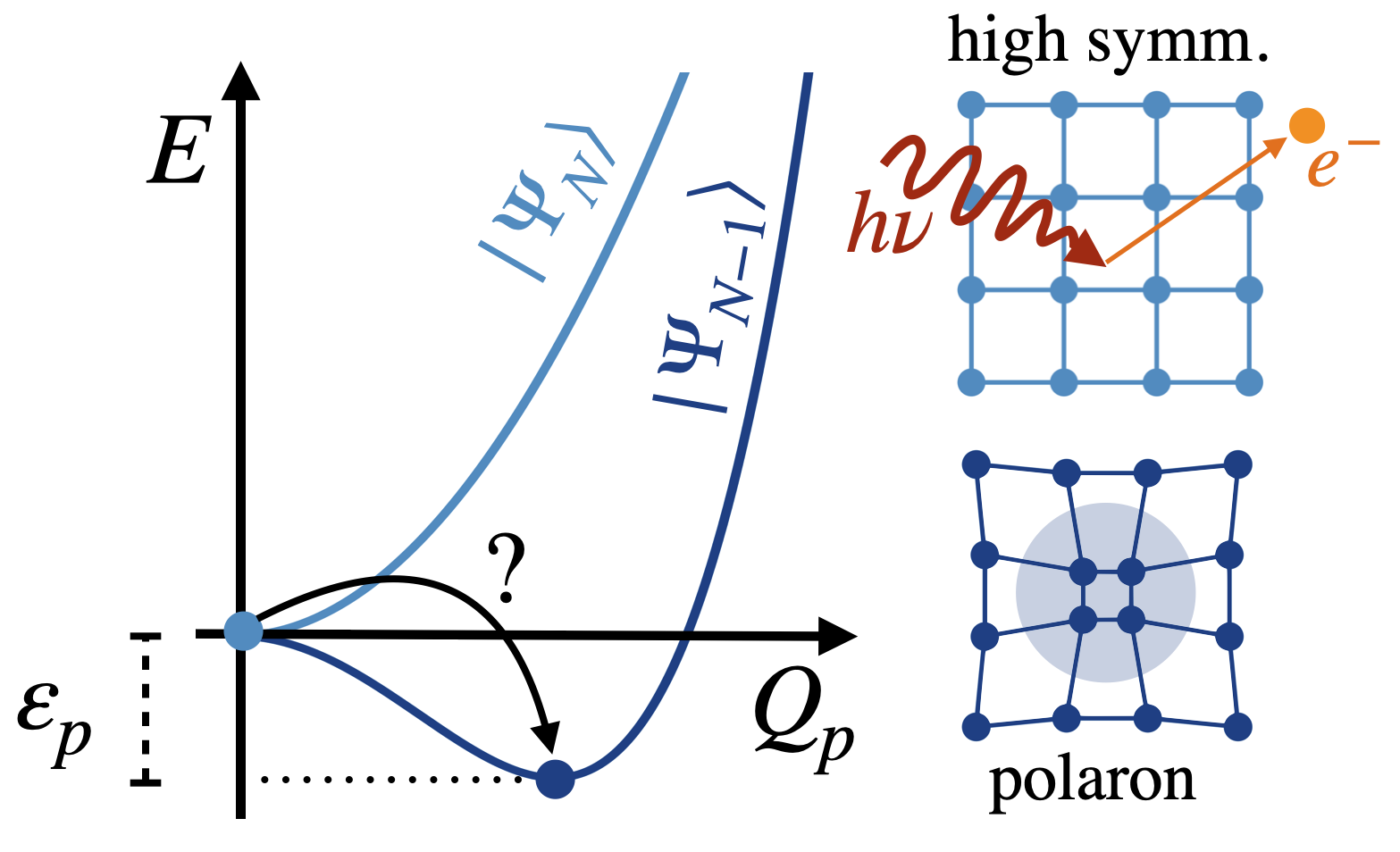}
\caption{Schematic illustration of the ground-state potential energy surface
$E$ for an $N$- (light) and $N-1$-electron systems (dark) as a function of
$Q_p$ -- a generalized coordinate connecting the high-symmetry and the
distorted polaronic structures. In the  $N-1$-electron system, polaron
formation proceed via a dynamical process in which a structural distortion
emerges in concomitance with carrier localization, resulting in an energy
lowering by the polaron formation energy $\varepsilon_p$.}\label{fig0}
\end{figure}

The advent of ultrafast spectroscopy \cite{boschini2024} and scattering
techniques \cite{filippetto_ultrafast_2022} is offering a powerful opportunity
to probe polaron formation in real time \cite{Miyata_2017,Li_2016,Seiler_2023},
enabling the direct observation of the emergence of symmetry-breaking lattice
distortions and carrier localization.  The dynamical process of polaron
formation, schematically illustrated in Fig.~\ref{fig0}, is initiated by the
introduction of an additional charge in the system -- for example, via
ultrafast charge transfer, photoabsorption, or threshold ionization
\cite{Kim_2019,DeSio_2016}.  The transition from the delocalized high-symmetry
state to a polaronic distorted structure requires a non-trivial dynamical and
dissipative process, which entails (i) the conversion of potential energy into
kinetic energy, initiating coherent lattice motion, \cite{Wang_2023} and (ii)
the gradual loss of coherence through dissipation, ultimately stabilizing the
polaronic ground state \cite{Chen_2025}.  

Conclusive identification of polaron formation in pump-probe experiments,
however, remains challenging, owing to the difficulty of disentangling spectral
fingerprints of polaronic from other dynamical processes, such as coherent
phonons or electronic processes occurring on similar time and energy scales
\cite{Cinquanta_2019,Cannelli_2021}.  These considerations highlight the
fundamental challenge of identifying distinctive dynamical fingerprints of
polaron formation that may guide their systematic and conclusive identification
in pump-probe experiments.  While several theoretical and computational
approach have been introduced and applied to describe polarons from first
principles -- including, e.g., hybrid density functional theory (DFT)
\cite{Falletta2022,Kokott_2018}, perturbative many-body methods
\cite{verdi_origin_2017,lafuente-bartolome_unified_2022,lafuente_abinitio_2022},
self-consistent variational problems \cite{sio_polarons_2019,sio_ab_2019}, and
Monte Carlo methods \cite{Mishchenko_2000, Luo_2025} --, these approaches have
thus far treated polarons as a static ground-state problem, which is not
directly applicable to describe the polaron formation as a non-equilibrium
problem in real time.  

In this manuscript, we develop a quantum-kinetic approach to investigate the
real-time dynamics of polaron formation from first principles.  We apply it to
investigate the localization of electron and hole polarons in the polar
insulators MgO. Our parameter-free simulations enable to predictively infer the
characteristic timescales of polaron formation, its dynamical fingerprint,  as
well as the dissipative mechanisms at play.  Overall, the polarons dynamics is
accompanied by a electronic and structural oscillations with distinct soft
frequency components that are absent in the ordinary phonon spectrum.  This
work thus identifies the distinctive dynamical blueprint of polarons, that may
enable to conclusively distinguish them from other forms of structural motion
(e.g., coherent phonons) thus guiding their identification in pump probe
experiments.  This work fills a critical gap in the theoretical modelling of
the ultrafast polaron dynamics, and structural symmetry breaking in driven
solids. 

The structural distortions arising from the formation of a polaron are
described by the operator $\Delta \hat{\boldsymbol{\tau}}_{\k p} (t)$, which
represents the displacement of the $\k$-th atom in the $p$-th unit cell
relative to equilibrium. It can be expanded in a normal-mode basis as
\cite{GiustinoRMP}:
$\Delta  \hat{\boldsymbol{\tau}}_{\k p} (t) = -{2}{ N_p^{-1}} \sum_{\bq\nu}
e^{i\bq{\bf R}_p} ({\hbar}/{2M_\k \omega_{\bq\nu}})^{{1}/{2}} {\bf
e}^{\k}_{\bq\nu} \hat B_{\bq\nu}^*(t)$, 
where $M_\k$ is the atomic mass, $\omega_{\bq\nu}$ and ${\bf e}^{\k}_{\bq\nu}$
are the frequency and eigenvector of a phonon with momentum $\bq$ and mode
$\nu$, respectively, ${\bf R}_p$ is a crystal-lattice vector, and $N_p$ the
number of unit cells in Born-von-Karman periodic boundary conditions
\cite{GiustinoRMP}.  The expectation value $B_{\bq\nu}(t) = \langle \hat
B_{\bq\nu}(t) \rangle$ is adimensional and quantifies the polaronic
distortion of the lattice along the phonon $\bq\nu$ \cite{sio_ab_2019}. 

The electronic counterpart of the polaron, quantifying the degree of
localization of an additional charge in the system at time $t$, can be
expressed via a wave-function {\it ansatz} $\ket{\Psi(t)} = \ket{\Psi_{\rm el}}
\otimes \ket{\psi_p (t)}$, where $\ket{\Psi_{\rm el}}$ is the $N$-electron
ground-state wave function, treated as time independent henceforth, and
$\ket{\psi_p (t)}$ is the wave function of an extra electron or hole injected
in the system at time $t=0$.  This {\it ansatz} is appropriate for the hole
polaron in MgO investigated here since the system resides in the
strong-coupling regime \cite{lafuente_abinitio_2022}.  By introducing the
expansion in Bloch basis $ \ket{\psi_p (t)} = N_p^{-1/2} \sum_{n\bk}
A_{n\bk}(t) \ket{\psi_{n\bk}}$, -- where $\ket{\psi_{n\bk}}$ is the Bloch state
for band $n$ and crystal momentum $\bk$ -- the degree of localization of the
additional charge and its dynamics are fully specified by the set of complex
time-dependent coefficients $A_{n\bk}(t)$.  Based on these definitions, the
problem of polaron dynamics reduces to determining the time evolution of the
phonon and electron envelope functions, $B_{\bq\nu}(t)$ and $A_{n\bk}(t)$,
respectively. 

To describe the dynamics of polaron formation, we derive a set of quantum
kinetic equations for the time evolution of $\hat B_{\bq\nu}(t)$.  In
Heisenberg picture, it obeys the equation of motion: 
$\partial^2 \hat{B}_{\bq\nu}/ {\partial t^2} =-\hbar^{-2} [[ \hat B_{\bq\nu},
\hat H] , \hat H]$.
The lattice Hamiltonian $\hat H = \hat H_{\rm ph} + \hat H_{\rm eph}$ includes
the harmonic phonon Hamiltonian $\hat H_{\rm ph}$ and the electron-phonon
interaction $ \hat H_{\rm eph}$.  We show in the Supplemental Materials \cite{sup} that by (i)
evaluating the nested commutators and (ii) taking the expectation value of
fermionic operators within the adiabatic approximation using the polaron
wave-function {\it ansatz} $\ket{\Psi(t)}$, one arrives at the following set of
coupled equations: 
\begin{align}\label{eq:Tpol0a}
&\D_t^2 B_{\bq\nu} + \gamma_{\bq\nu} \D_t B_{\bq\nu} + \w_{\bq\nu}^2 B_{\bq\nu} = \\
&
\quad \quad \quad \quad 
\quad \quad \quad \quad 
\frac{\omega_{\bq \nu}}{ \hbar N_p } \sum_{m n \bk} g_{mn}^\nu(\bk,\bq)
A_{m \bk+\bq}^* A_{n \bk}\quad, \nonumber \\
&[\ve_{n\bk} -\ve] A_{n\bk} =  N_p^{-1}  \sum_{m} \sum_{\bq\nu}
[g_{mn}^\nu (\bk,\bq)]^* A_{m\bk+\bq}   B_{\bq \nu},
\label{eq:Tpol0b}
\end{align}
where $\partial_t = \partial/\partial t$ and $\partial_t^2 =
\partial^2/\partial t^2$, and $\gamma_{\bq\nu}$ denotes the decoherence rate
introduced according to Ref.~\cite{Pan2025}.  $g_{nm}^\nu(\bk,\bq)$ is the
electron-phonon coupling matrix element and $\ve$ is polaron eigenvalue.
Equation~\eqref{eq:Tpol0b} is derived through variational minimization of the
total electronic energy at time $t$ with the constrained of preserving
normalization ($N_p^{-1}\sum_{n\bk} | A_{n\bk}|^2 = 1$)
\cite{sio_polarons_2019}.  This step  corresponds to treat the electronic
response to a time-dependent structural distortion within the adiabatic
approximation.  Equations~\eqref{eq:Tpol0a} and \eqref{eq:Tpol0b} are the {\it
time-dependent polaron equations} that can be solved by time propagation to
infer the dynamics of the electron and phonon envelope functions, $ A_{n
\bk}(t)$ and  $B_{\bq\nu} (t)$ throughout the polaron formation.  A detailed
derivation of this result is reported in the Supplemental Materials \cite{sup}.  

We implemented Eqs.~\eqref{eq:Tpol0a} and \eqref{eq:Tpol0b} in a modified
version of the {\tt EPW } code \cite{lee_electronphonon_2023} and applied them
to investigate the polaron dynamics of the prototypical polar semiconductor
MgO, for which polaron formation is well characterized in the ground state
\cite{Kokott_2018, Falletta2022, Chae_2025}.  Electronic and vibrational
properties are obtained from density-functional theory (DFT) and density
functional perturbation theory calculations (DFPT) \cite{BaroniRMP} as
implemented in the plane-wave pseudo-potential code {\tt Quantum Espresso}
\cite{giannozzi2017advanced}.  Equations~\eqref{eq:Tpol0a} and
\eqref{eq:Tpol0b} are solved on a homogeneous 24$\times$24$\times$24 grid for
$\bk$ and $\bq$ via second-order Runge-Kutta time stepping using a time
stepping of 1~fs for a total simulation time of 4~ps.  Further details on the
numerical implementation and computational parameters are reported in the
Supplemental Materials~\cite{sup}. 

\begin{figure*}[t]
\centering
\includegraphics[width=0.98\linewidth]{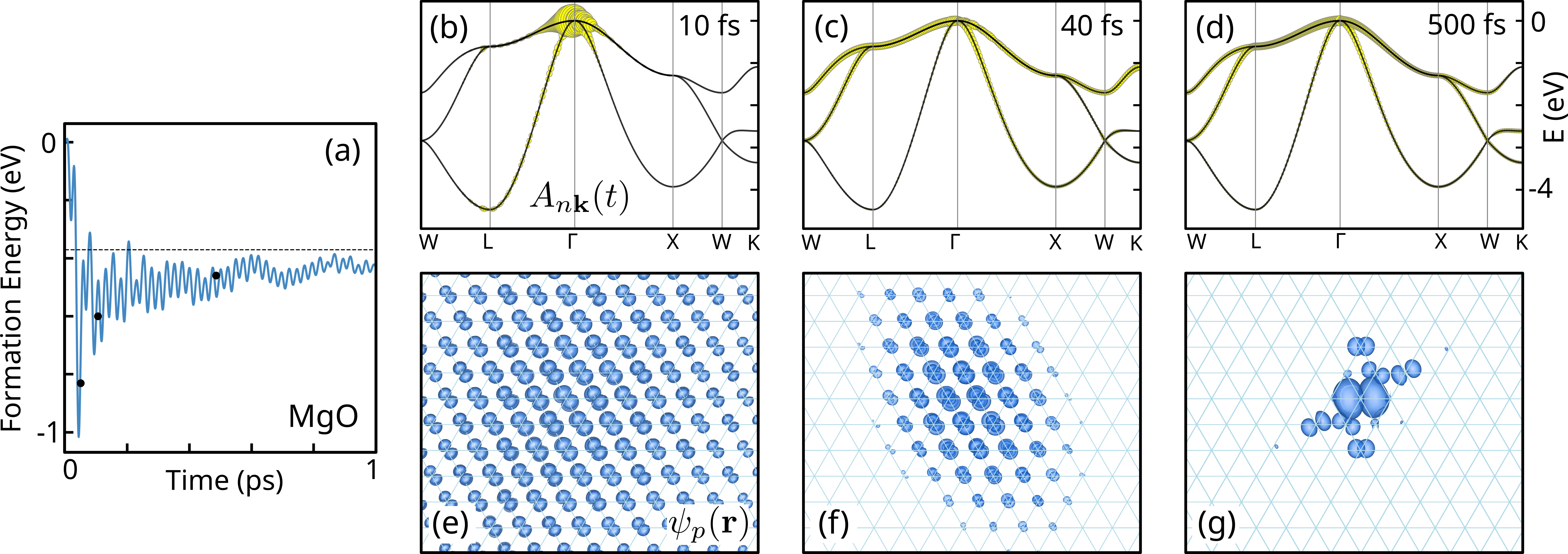}
\caption{ (a) Time dependence of the polaron formation energy throughout the
dynamical formation of a hole polaron in MgO.  The static polaron formation
energy obtained from the solution of the static polaron equation or
Ref.~\cite{sio_ab_2019} is reported as dashed horizontal line.  (b-d) Hole
envelope function $A_{n\bk}(t)$ obtained from the time propagation of the
time-dependent polaron equations  for time delays of $t=10$, 40, and 500~fs,
respectively (marked by black dots in panel (a)).  (e-g) Density of an extra
hole injected at the top of conduction band for the same time steps of panels
(b-d).  }\label{fig1}
\end{figure*}
 
The polaron formation energy $E_p = \varepsilon + N_p^{-1}\sum_{
\bq\nu}|B_{\bq\nu}|^2 \hbar\w_{\bq\nu}$ quantifies the energy gained by the crystal by forming a polaron following injection of a charge carrier.  The time-dependent formation energy obtained from the solution of Eqs.~\eqref{eq:Tpol0a} and \eqref{eq:Tpol0b} for MgO is reported in
Fig.~\ref{fig1}~(a) for time delays up to 1~ps following charge injection.  At time $t=0$, the extra charge is delocalized through the crystal and the structure undistorted, thus $E_p=0$. For $t>0$, $E_p$ exhibits damped oscillations over timescales dictated by  the longitudinal optical (LO) phonon period, indicating an oscillatory structural dynamics that accompanies the emergence of a polaron.  Phonon-phonon interaction is the primary mechanisms for phonon decoherence in insulators \cite{Pan2025}, and it causes the damping of the oscillations within few picoseconds.  After damping, we find $E_p(t=5~{\rm ps}) \simeq -0.4$~eV, which coincides with the static formation energy $E_p^{\rm static}$ \cite{Chae_2025} (marked by a dashed line in Fig.~\ref{fig1}~(a)) obtained from the ground-state formalism developed in Refs.~\cite{sio_ab_2019,sio_polarons_2019}. Supplementary Fig.~S1 further illustrates that static and dynamical formation energies agree for all choices of the ${\bk}$-point meshes. This agreement demonstrates that once the polaron is fully formed, the dynamical results of Eqs.~\eqref{eq:Tpol0a} and \eqref{eq:Tpol0b} converge to the established ground-state theory, thereby providing a consistency check that validates our simulations. 
To validate our approach and its numerical implementation, we conducted calculations of polaron formation within the DFT supercell formalism, with self-interaction 
correction based on Refs.~\textcite{Sadigh2015} and \textcite{Dai2025}. 
Specifically, we considered a 10$\times$10$\times$10 MgO supercell consisting of 2000 atoms, and compared polaron formation energy and electron densities at various timesteps along the non-equilibrium trajectories (Figs.~S2 and S3 \cite{sup}). Overall, we find good agreement between the time-dependent polaron equations and DFT supercell calculations, confirming the correctness of the proposed framework.  
 
To inspect the dynamics of charge localization, we report in
Figs.~\ref{fig1}~(b-d) the envelope function $A_{n\bk}$ for time delays of $t=10$, 40, and 500~fs, along a high-symmetry path in the Brillouin zone.
Initially, $A_{n\bk}$ is concentrated in the vicinity of the $\Gamma$
high-symmetry point reflecting the delocalized character of the extra charge in real space.  On longer times, the envelope function progressively spread out to encompass different momenta in the Brillouin zone, indicating the increased localization of the polaron charge density in real space.  The localization process is further revealed by the isosurface plots of the polaron wave function $\psi_p(\br)$ -- reported in Figs.\ref{fig1}~(e–f) at $t=10$, 40, and 500~fs --, which indicate the emergence of distinct intermediate states of electronic localization during polaron formation.
The full time dependence envelope function $A_{n\bk}$ and of the polaron density is illustrated in the Supplementary Movies 1 and 2 \cite{sup}. 
In the SM, we further present results for the  dynamical formation of large electron polarons in LiF (see supplementary discussion Section VII and Fig. S8 in \cite{sup}). In LiF, the polaron extends over a larger region of the lattice compared to the hole polaron in MgO, illustrating how the framework is capable of describing not only strongly localized polarons but also larger and more delocalized ones. 

\begin{figure*}[t]
\centering
\includegraphics[width=0.98\linewidth]{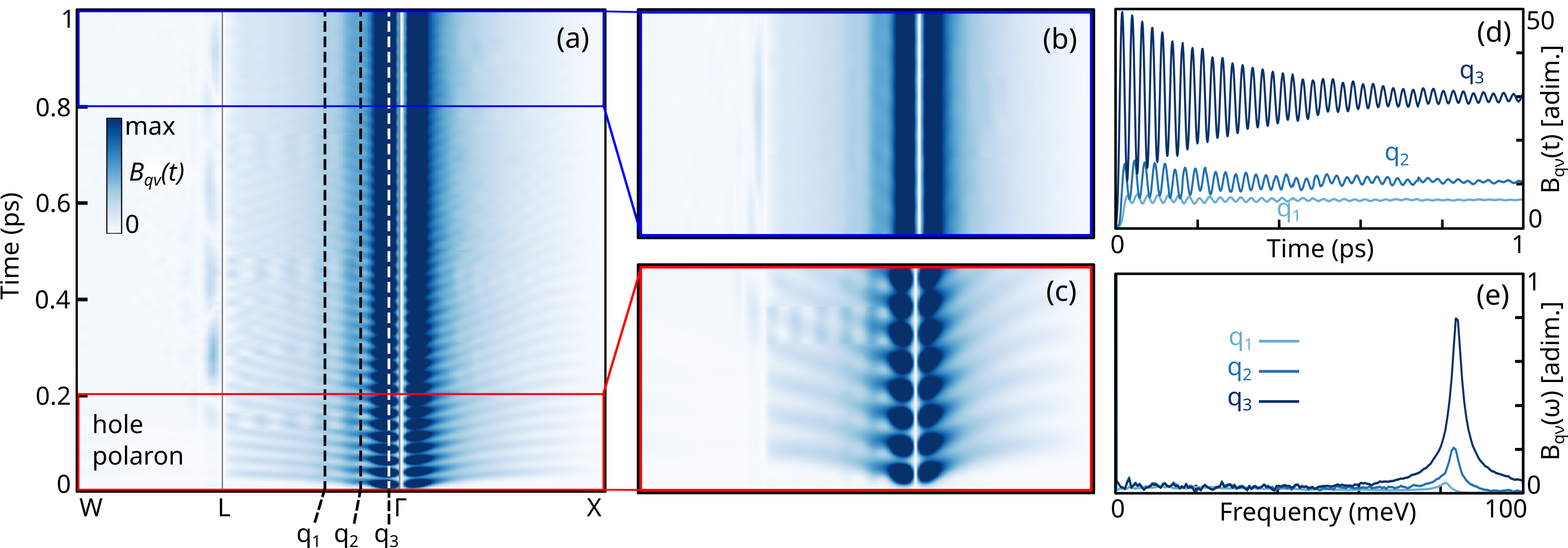}
\caption{(a) Ultrafast modulation of the phonon polaron envelope function
$B_{\bq \nu}(t)$ during the  formation of a hole polaron in MgO.  The phonon mode index $\nu$ is set to the longitudinal optical phonon, and the momentum $\bq$ runs along the W-L-$\Gamma$-X high-symmetry path in the Brillouin zone. (b-c) Enlarged views of the regions highlighted by the blue and red rectangles in panel (a).  (d) Time dependence of the phonon envelope function $B_{\bq \nu}(t)$ for the momenta $q_1,q_2,q_3$ marked by vertical lines in panel~(a). (e) Fourier transform  $B_{\bq\nu}(\w)$ of the data in (d).}\label{fig2}
\end{figure*}

To illustrate the dynamics of the lattice distortion accompanying the polaron dynamics in MgO, we examine the time evolution of the phonon polaron envelope function  $B_{\bq\nu}(t)$, focusing on the LO mode ($\nu={\rm LO}$) and wave vectors $\bq$ along the  W-L-$\Gamma$-X high-symmetry path in  reciprocal space. $B_{\bq\nu}(t)$ quantifies the displacement of the lattice along the phonon $\bq\nu$ at time $t$, and it is thus a direct indicator of the structural motion.  Its dynamics -- illustrated in Fig.~\ref{fig2}~(a) for the formation of a hole polarons and times up to 1~ps -- exhibits pronounced oscillations across the entire Brillouin zone. The oscillation period varies significantly with momentum, increasing progressively as one moves away from the $\Gamma$ point (Fig.~\ref{fig2}~(c)).  Wave vectors with larger oscillation amplitudes (darker color in Fig.~\ref{fig2}~(a)) correspond to modes contributing more significantly to the overall polaronic distortion.
Oscillations are damped within 1~ps (Fig.~\ref{fig2}~(b)), suggesting that the polaron has formed within this time interval.  A quantitative definition of the polaron formation time can be formulated by requiring that 90\% of the atomics displacements for all atoms in the Born-von-Karman supercell and relative to the final polaron structure fall below a threshold value of $\Delta_{\rm th} = 0.005$~\AA. Based on this criterion, we estimate the formation time for hole polarons in MgO to 0.8~ps (see Supplementary discussion V and Figs.~S4 and S5 in \cite{sup}).  This value agrees well with the characteristic timescales of polaron formation estimated via pump-probe optical experiments \cite{Chen_2025}. 
{While no experimental data for dynamical polaron formation in MgO are available, existing time-resolved experiments for double perovskites \cite{Wu_2021,Bretschneider_2018,Wright_2021} 
report polaron formation times ranging between 1 and 5~ps \cite{Bretschneider_2018}, matching closely the ones reported in our work.}  
Overall, the timescale of polaron formation emerges as an intrinsic property of the material, determined by (i) the specific pathway connecting the distorted and undistorted structures, and (ii) the characteristic timescales of decoherence arising from phonon-phonon
interactions. Conversely, the LO phonon period  ($\sim$50~fs in MgO) does not play a significant role on these timescales. To further clarify this point, we derive an approximate analytical formula to  estimate the polaron formation time $t_{\rm f }$ (see Supplementary discussion VI \cite{sup}). For low damping ($\gamma\ll\w$), the formation time can be expressed as $t_{\rm f} = \gamma^{-1}\ln(\tau_{\rm avg}/\Delta_{\rm th})$, where $\tau_{\rm avg}$ is the average (absolute) atomic displacement induced by the polaron,  $\Delta_{\rm th}$ the displacement threshold, and $\gamma$ the average damping rate. This expression further emphasizes that the polaron formation times are mostly governed by the damping rate $\gamma$ and by the magnitude of the displacement associated with the formation of a polaron.  Using parameters for MgO, we obtain $t_{\rm f} = 0.7$~ps, in agreement with the ab-initio value of 0.8~ps.

\begin{figure*}[t]
\centering
\includegraphics[width=0.98\linewidth]{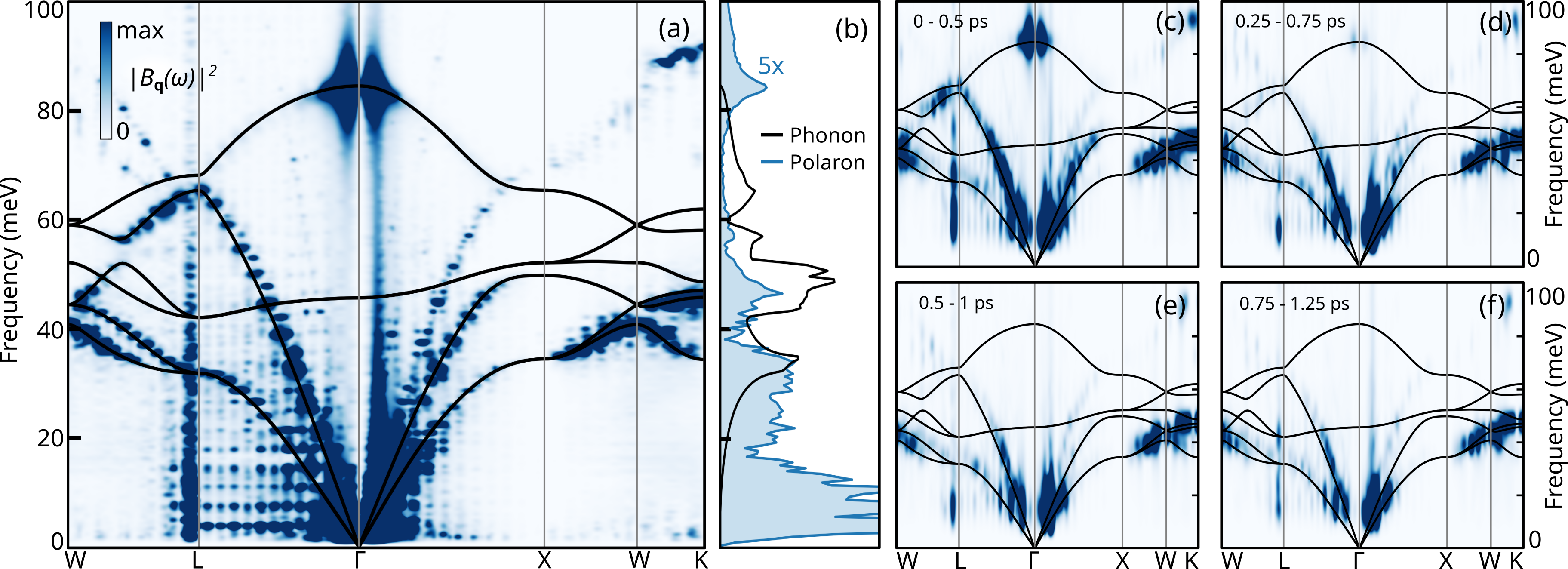}
\caption{(a) Polaron spectral function $\mathcal{B}^2_{\bq} (\omega) = \sum_\nu
|B_{\bq \nu}(\omega)|^2$ for momenta along the W-L-$\Gamma$-X high-symmetry
path. The phonon dispersion obtained from density-functional perturbation
theory (black line) is included for comparison.  (b) Comparison between the
polaron (blue) and the phonon density of state (black).  (c-f) Polaron spectral
function  $\mathcal {B}^2_{\bq} (\omega)$ obtained by restricting the Fourier
transforms to time intervals of 0.5~ps for different time delays after the
onset of the dynamics.  }\label{fig3}
\end{figure*}

The time dependence of $B_{\bq\nu}(t)$ for the LO phonon is further illustrated
in Fig.~\ref{fig2}~(d) for the wave vectors $q_1$, $q_2$, and $q_3$ (vertical
lines in Fig.~\ref{fig2}~(c)), revealing a structural dynamics characterized by
damped oscillations. The Fourier transform of the time series -- defined as
$B_{\bq\nu}(\w) = \int dt B_{\bq\nu}(t)e^{i\w t}$ and  shown in
Fig.~\ref{fig2}~(e) -- yields frequencies in the vicinity of the LO phonon
frequencies, thus,  identifying this mode with the dominant mechanism behind
the fast modulation of the structural distortion.  In addition to the LO
phonons, however, other frequencies arise that cannot be directly attributed to
LO vibrations.  {To analyze the entire frequency spectrum of the structural
oscillations,  we define the polaron spectral function as ${\mathcal B}^2_{\bq}
(\omega) = \sum_\nu  |B_{\bq\nu}  (\w)|^2$.  This quantity serves as a global
indicator of the vibrational spectrum  arising throughout polaron formation,
and it can be directly related to the number of phonons contributing to the
polaron \cite{sio_ab_2019}.} The polaron spectral function ${\mathcal
B}^2_{\bq} (\omega)$ is illustrated in Fig.~\ref{fig3}~(a) as a color map as a
function of momentum and frequency, while its mode-resolved components
$|B_{\bq\nu}  (\w)|^2$ are shown in Fig.~S6 \cite{sup}.  The phonon dispersion obtained from DFPT is superimposed for comparison.  Here, dark regions indicate the
presence of structural oscillations at specific frequencies and momenta,
whereas white denotes the absence of oscillatory motion. Oscillations are
confined to a well-defined region of momentum space corresponding to the modes
actively involved in polaron formation.  Most importantly, the frequency
spectrum of the polaron differ substantially from the phonon spectrum.  In
particular, we observe a strong contribution of low-frequency components that
have no counterpart in the phonon dispersion.  These findings suggest that
polaron formation is characterized by distinctive dynamical signatures, which
could serve as markers for unambiguously identifying polarons in pump–probe
experiments.  The difference between the vibrational frequency spectrum of
phonons and polarons is further highlighted in Fig.\ref{fig3}~(b), where the
polaron density of states (DOS), defined as $\mathcal{D}_{\rm p} (\omega) =
\int_{\Omega_{\rm BZ}} {d\bq} \mathcal{B}^2_{\bq} (\omega)$, is compared to the
phonon DOS. The decomposition of $\mathcal{D}_{\rm p} (\omega) $  in its
mode-resolved contributions is reported in Fig.~S7 \cite{sup}.  To disentangle
the timescales at which different frequency components emerge in the dynamics,
Figures~\ref{fig3}~(c-f) reports the polaron spectral function
$\mathcal{B}^2_{\bq} (\omega) $ obtained by restricting the Fourier transform
to time intervals of $500$~fs. This analysis reveals that structural
oscillations at the LO frequency only persist throughout the initial phase of
polaron formation. In contrast, low-frequency oscillations persist on longer
time scales.  We attribute these modes to the distinctive vibration dynamics of
the polaron that arise from the long-range structural reorganization of the
crystal,  giving rise to soft vibrational motion absent in the equilibrium
cubic structure. 

These results offer a guideline for identifying the dynamical fingerprints of
polaron formation in pump-probe experiments.  In ultrafast diffuse scattering
\cite{filippetto_ultrafast_2022}, polarons lead to an enhanced scattering
intensity in the vicinity of the Bragg peaks
\cite{britt_momentum-resolved_2024}.  The time-dependent oscillations of the
polaron envelop function $B_{\bq\nu}(t)$ are expected to induce a coherent
modulation of this diffuse signal, with a characteristic frequency spectrum as
shown in Fig.~\ref{fig3}~(a).  Importantly, this behavior distinguishes
polarons from coherent phonons, which affect only the Bragg peak intensities
without generating corresponding features in the diffuse scattering background.
The frequency spectrum associated with polaron formation should further leave
distinctive signatures in time-resolved optical and ARPES measurements. In
presence of polaronic distortions, the band energy $\ve_{n\bk}$ is renormalized
to $\ve_{n\bk} +\Delta\ve_{n\bk}(t)$, where  $\Delta \ve_{n\bk} (t)=
N_p^{-1}\sum_{\bq\nu} g_{nn}^\nu(\bk,\bq) B_{\bq\nu}(t)$
\cite{lafuente-bartolome_unified_2022,emeis_coherent_2025}.  Polaron formation
can thus be inferred from time-dependent oscillations in the optical or
photoemission signal. In particular, the emergence of low-frequency components,
typically not associated with coherent phonons, may serve as a distinguishing
feature of polaron dynamics, providing a means to separate them from other
forms of coherent lattice motion.  These considerations provide a rigorous
criteria for distinguishing polaron formation from other non-equilibrium
phenomena in time-resolved experiments.

In conclusion, we developed a first-principles theoretical framework to
describe the ultrafast dynamics of polaron localization.  We applied it to
investigate the formation of polarons in real time in the polar insulator MgO.
Besides estimating the characteristic timescales of polaron localization and
decoherence, our simulations reveal a nontrivial dynamical behavior,
characterized by damped structural oscillations and a frequency spectrum
exhibiting distinctive features beyond those of ordinary harmonic lattice
vibrations. These results demonstrate that polaron formation is accompanied by
characteristic dynamical fingerprints, which can serve as markers to
distinguish polarons from other electronic and vibrational excitations. This
insight provides a valuable foundation for interpreting time-resolved
pump-probe experiments and for advancing our understanding of nonequilibrium
quasiparticle dynamics in polar crystals.

\acknowledgments
This work was funded by the European Union as part of the MSCA Doctoral Network
TIMES (Grant Agreement No. 101118915) and by the Deutsche
Forschungsgemeinschaft (DFG), Projects No. 499426961 and 443988403. The authors
gratefully acknowledge the computing time provided by the high-performance
computer Lichtenberg at the NHR Centers NHR4CES at TU Darmstadt (Project
p0021280). F.C. acknowledges the J.~Tinsley Oden Fellowship Research Program of
the Oden Institute at the University of Texas Austin. F.G. was supported by the
Computational Materials Sciences Program funded by the US Department of Energy,
Office of Science, Basic Energy Sciences, under award no. DE-SC0020129.
J.L.-B. was supported by Grant No. IT-1527-22, funded by the Department of
Educa- tion, Universities and Research of the Basque Govern- ment, and Grant
no. PID2022-137685NB-I00, funded by MCIN/AEI/10.13039/501100011033/ and by
“ERDF A way of making Europe”.

\bibliography{references}

@misc{Pan2025,
      title={Origin of phonon decoherence}, 
      author={Yiming Pan and Christoph Emeis and Stephan Jauernik and Michael Bauer and Fabio Caruso},
      year={2025},
      eprint={2502.01529},
      archivePrefix={arXiv},
      url={https://arxiv.org/abs/2502.01529}, 
}

@article{lee_electronphonon_2023,
	title = {Electron–phonon physics from first principles using the {EPW} code},
	volume = {9},
	issn = {2057-3960},
	url = {https://www.nature.com/articles/s41524-023-01107-3},
	doi = {10.1038/s41524-023-01107-3},
	number = {1},
	urldate = {2024-07-23},
	journal = {npj Comput. Mater.},
	author = {Lee, Hyungjun and Poncé, Samuel and Bushick, Kyle and Hajinazar, Samad and Lafuente-Bartolome, Jon and Leveillee, Joshua and Lian, Chao and Lihm, Jae-Mo and Macheda, Francesco and Mori, Hitoshi and Paudyal, Hari and Sio, Weng Hong and Tiwari, Sabyasachi and Zacharias, Marios and Zhang, Xiao and Bonini, Nicola and Kioupakis, Emmanouil and Margine, Elena R. and Giustino, Feliciano},
	month = aug,
	year = {2023},
	pages = {156},
}

@article{franchini_polarons_2021,
	title = {Polarons in materials},
	volume = {6},
	issn = {2058-8437},
	url = {https://www.nature.com/articles/s41578-021-00289-w},
	doi = {10.1038/s41578-021-00289-w},
	number = {7},
	urldate = {2025-05-09},
	journal = {Nat. Rev. Mater.},
	author = {Franchini, Cesare and Reticcioli, Michele and Setvin, Martin and Diebold, Ulrike},
	month = mar,
	year = {2021},
	pages = {560--586},
}

@article{verdi_origin_2017,
	title = {Origin of the crossover from polarons to {Fermi} liquids in transition metal oxides},
	volume = {8},
	issn = {2041-1723},
	url = {http://www.nature.com/articles/ncomms15769},
	doi = {10.1038/ncomms15769},
	number = {1},
	urldate = {2021-09-24},
	journal = {Nat. Commun.},
	author = {Verdi, Carla and Caruso, Fabio and Giustino, Feliciano},
	month = aug,
	year = {2017},
	pages = {15769},
}

@article{sio_polarons_2019,
        title = {Polarons from First principles, without Supercells},
        volume = {122},
        issn = {0031-9007, 1079-7114},
        url = {https://link.aps.org/doi/10.1103/PhysRevLett.122.246403},
        doi = {10.1103/PhysRevLett.122.246403},
        number = {24},
        urldate = {2022-09-23},
        journal = {Phys. Rev. Lett.},
        author = {Sio, Weng Hong and Verdi, Carla and Poncé, Samuel and Giustino, Feliciano},
        month = jun,
        year = {2019},
        pages = {246403},
}

@article{sio_ab_2019,
        title = {{{Ab} initio} theory of polarons: {Formalism} and applications},
        volume = {99},
        issn = {2469-9950, 2469-9969},
        shorttitle = {{{Ab} initio} theory of polarons},
        url = {https://link.aps.org/doi/10.1103/PhysRevB.99.235139},
        doi = {10.1103/PhysRevB.99.235139},
        number = {23},
        urldate = {2022-09-23},
        journal = {Phys. Rev. B},
        author = {Sio, Weng Hong and Verdi, Carla and Poncé, Samuel and Giustino, Feliciano},
        month = jun,
        year = {2019},
        pages = {235139},
}

@article{GiustinoRMP,
  title = {Electron-phonon interactions from first principles},
  author = {Giustino, Feliciano},
  journal = {Rev. Mod. Phys.},
  volume = {89},
  issue = {1},
  pages = {015003},
  numpages = {63},
  year = {2017},
  month = {Feb},
  publisher = {American Physical Society},
  doi = {10.1103/RevModPhys.89.015003},
  url = {https://link.aps.org/doi/10.1103/RevModPhys.89.015003}
}

@article{britt_momentum-resolved_2024,
	title = {A momentum-resolved view of polaron formation in materials},
	volume = {10},
	issn = {2057-3960},
	url = {https://www.nature.com/articles/s41524-024-01347-x},
	doi = {10.1038/s41524-024-01347-x},
	number = {1},
	urldate = {2024-08-28},
	journal = {npj Comput. Mater.},
	author = {Britt, Tristan L. and Caruso, Fabio and Siwick, Bradley J.},
	month = aug,
	year = {2024},
	pages = {178},
}

@article{filippetto_ultrafast_2022,
	title = {Ultrafast electron diffraction: {Visualizing} dynamic states of matter},
	volume = {94},
	issn = {0034-6861, 1539-0756},
	shorttitle = {Ultrafast electron diffraction},
	url = {https://link.aps.org/doi/10.1103/RevModPhys.94.045004},
	doi = {10.1103/RevModPhys.94.045004},
	number = {4},
	urldate = {2023-07-03},
	journal = {Rev. Mod. Phys.},
	author = {Filippetto, D. and Musumeci, P. and Li, R. K. and Siwick, B. J. and Otto, M. R. and Centurion, M. and Nunes, J. P. F.},
	month = dec,
	year = {2022},
	pages = {045004},
}

@article{emeis_coherent_2025,
	title = {Coherent {Phonons} and {Quasiparticle} {Renormalization} in {Semimetals} from {First} {Principles}},
	volume = {15},
	issn = {2160-3308},
	url = {https://link.aps.org/doi/10.1103/PhysRevX.15.021039},
	doi = {10.1103/PhysRevX.15.021039},
	number = {2},
	urldate = {2025-05-11},
	journal = {Phys. Rev. X},
	author = {Emeis, Christoph and Jauernik, Stephan and Dahiya, Sunil and Pan, Yiming and Jensen, Carl E. and Hein, Petra and Bauer, Michael and Caruso, Fabio},
	month = may,
	year = {2025},
	pages = {021039},
}

@article{lafuente-bartolome_unified_2022,
	title = {Unified Approach to Polarons and Phonon-Induced Band Structure Renormalization},
	volume = {129},
	issn = {0031-9007, 1079-7114},
	url = {https://link.aps.org/doi/10.1103/PhysRevLett.129.076402},
	doi = {10.1103/PhysRevLett.129.076402},
	number = {7},
	urldate = {2022-09-23},
	journal = {Phys. Rev. Lett.},
	author = {Lafuente-Bartolome, Jon and Lian, Chao and Sio, Weng Hong and Gurtubay, Idoia G. and Eiguren, Asier and Giustino, Feliciano},
	month = aug,
	year = {2022},
	pages = {076402},
}

@article{Falletta2022,
  title = {Many-Body Self-Interaction and Polarons},
  author = {Falletta, Stefano and Pasquarello, Alfredo},
  journal = {Phys. Rev. Lett.},
  volume = {129},
  issue = {12},
  pages = {126401},
  numpages = {6},
  year = {2022},
  month = {Sep},
  publisher = {American Physical Society},
  doi = {10.1103/PhysRevLett.129.126401},
  url = {https://link.aps.org/doi/10.1103/PhysRevLett.129.126401}
}

@article{Kokott_2018,
doi = {10.1088/1367-2630/aaaf44},
url = {https://dx.doi.org/10.1088/1367-2630/aaaf44},
year = {2018},
month = {mar},
publisher = {IOP Publishing},
volume = {20},
number = {3},
pages = {033023},
author = {Kokott, Sebastian and Levchenko, Sergey V and Rinke, Patrick and Scheffler, Matthias},
title = {First-principles supercell calculations of small polarons with proper account for long-range polarization effects},
journal = {New J. Phys.},
}

@Article{Zhang_2007,
author={Zhang, S. X.
and Kundaliya, D. C.
and Yu, W.
and Dhar, S.
and Young, S. Y.
and Salamanca-Riba, L. G.
and Ogale, S. B.
and Vispute, R. D.
and Venkatesan, T.},
title={Niobium doped {$\rm TiO_2$}: Intrinsic transparent metallic anatase versus highly resistive rutile phase},
journal={J. Appl. Phys.},
year={2007},
month={Jul},
day={02},
volume={102},
number={1},
pages={013701},
issn={0021-8979},
doi={10.1063/1.2750407},
url={https://doi.org/10.1063/1.2750407}
}

@article{Wright_2021,
author = {Wright, Adam D. and Buizza, Leonardo R. V. and Savill, Kimberley J. and Longo, Giulia and Snaith, Henry J. and Johnston, Michael B. and Herz, Laura M.},
title = {Ultrafast Excited-State Localization in {$\rm Cs_2AgBiBr_6$} Double Perovskite},
journal = {J. Phys. Chem. Lett.},
volume = {12},
number = {13},
pages = {3352-3360},
year = {2021},
doi = {10.1021/acs.jpclett.1c00653},
URL = {https://doi.org/10.1021/acs.jpclett.1c00653}
}

@article{Redondo_2024,
author = {Jesus Redondo  and Michele Reticcioli  and Vit Gabriel  and Dominik Wrana  and Florian Ellinger  and Michele Riva  and Giada Franceschi  and Erik Rheinfrank  and Igor Sokolović  and Zdenek Jakub  and Florian Kraushofer  and Aji Alexander  and Eduard Belas  and Laerte L. Patera  and Jascha Repp  and Michael Schmid  and Ulrike Diebold  and Gareth S. Parkinson  and Cesare Franchini  and Pavel Kocan  and Martin Setvin },
title = {Real-space investigation of polarons in hematite {$\rm Fe_2O_3$}},
journal = {Sci. Adv.},
volume = {10},
number = {44},
pages = {eadp7833},
year = {2024},
doi = {10.1126/sciadv.adp7833},
URL = {https://www.science.org/doi/abs/10.1126/sciadv.adp7833},
}

@article{Luo_2018,
author={Luo, Jiajun
and Wang, Xiaoming
and Li, Shunran
and Liu, Jing
and Guo, Yueming
and Niu, Guangda
and Yao, Li
and Fu, Yuhao
and Gao, Liang
and Dong, Qingshun
and Zhao, Chunyi
and Leng, Meiying
and Ma, Fusheng
and Liang, Wenxi
and Wang, Liduo
and Jin, Shengye
and Han, Junbo
and Zhang, Lijun
and Etheridge, Joanne
and Wang, Jianbo
and Yan, Yanfa
and Sargent, Edward H.
and Tang, Jiang},
title={Efficient and stable emission of warm-white light from lead-free halide double perovskites},
journal={Nature},
year={2018},
month={Nov},
day={01},
volume={563},
number={7732},
pages={541-545},
issn={1476-4687},
doi={10.1038/s41586-018-0691-0},
url={https://doi.org/10.1038/s41586-018-0691-0}
}

@article{Mishchenko_2000,
  title = {Diagrammatic quantum {Monte} {Carlo} study of the {Fr\"ohlich} polaron},
  author = {Mishchenko, A. S. and Prokof'ev, N. V. and Sakamoto, A. and Svistunov, B. V.},
  journal = {Phys. Rev. B},
  volume = {62},
  issue = {10},
  pages = {6317--6336},
  numpages = {0},
  year = {2000},
  month = {Sep},
  publisher = {American Physical Society},
  doi = {10.1103/PhysRevB.62.6317},
  url = {https://link.aps.org/doi/10.1103/PhysRevB.62.6317}
}

@misc{Chae_2025,
      title={Extreme-Band-Gap Semiconductors with Shallow Dopants and Mobile Carriers}, 
      author={Sieun Chae and Nocona Sanders and Kelsey A. Mengle and Amanda Wang and Xiao Zhang and Jon Lafuente Bartolome and Kaifa Luo and Yen-Chun Huang and Feliciano Giustino and John T. Heron and Emmanouil Kioupakis},
      year={2025},
      eprint={2506.07284},
      archivePrefix={arXiv},
      url={https://arxiv.org/abs/2506.07284}, 
}

@article{Luo_2025,
  author={Luo, Yao and Park, Jinsoo and Bernardi, Marco},
  title={First-principles diagrammatic Monte Carlo for electron--phonon interactions and polaron},
  journal={Nature Physics},
  year={2025},
  month={Jul},
  day={10},
  issn={1745-2481},
  doi={10.1038/s41567-025-02954-1},
  url={https://doi.org/10.1038/s41567-025-02954-1}
}

@article{boschini2024,
  title = {Time-resolved {ARPES} studies of quantum materials},
  author = {Boschini, Fabio and Zonno, Marta and Damascelli, Andrea},
  journal = {Rev. Mod. Phys.},
  volume = {96},
  issue = {1},
  pages = {015003},
  numpages = {56},
  year = {2024},
  month = {Feb},
  publisher = {American Physical Society},
  doi = {10.1103/RevModPhys.96.015003},
  url = {https://link.aps.org/doi/10.1103/RevModPhys.96.015003}
}

@article{Liu2023,
  author       = {Liu, Huiru and Wang, Aolei and Zhang, Ping and Ma, Chen and Chen, Caiyun and Liu, Zijia and Zhang, Yi‑Qi and Feng, Baojie and Cheng, Peng and Zhao, Jin and Chen, Lan and Wu, Kehui},
  title        = {Atomic‐scale manipulation of single‐polaron in a two‐dimensional semiconductor},
  journal      = {Nat. Commun.},
  volume       = {14},
  number       = {1},
  pages        = {3690},
  year         = {2023},
  doi          = {10.1038/s41467-023-39361-0},
  publisher    = {Nature Publishing Group},
  url          = {https://doi.org/10.1038/s41467-023-39361-0}
}

@article{Miyata_2017,
  author  = {Miyata, Kiyoshi and Meggiolaro, Daniele and Trinh, M.~Tuan
             and Joshi, Prakriti P. and Mosconi, Edoardo and Jones, Skyler C.
             and De Angelis, Filippo and Zhu, X.-Y.},
  title   = {Large Polarons in Lead Halide Perovskites},
  journal = {Sci. Adv.},
  year    = {2017},
  volume  = {3},
  number  = {8},
  pages   = {e1701217},
  doi     = {10.1126/sciadv.1701217}
}

@article{Li_2016,
  author  = {Li, Junjie and Yin, Wei-Guo and Wu, Lijun and Zhu, Pengfei
             and Konstantinova, Tatiana and Tao, Jing and Yang, Junjie
             and Cheong, Sang-Wook and Carbone, Fabrizio and Misewich, James~A.
             and Hill, John~P. and Wang, Xijie and Cava, Robert~J. and Zhu, Yimei},
  title   = {Dichotomy in Ultrafast Atomic Dynamics as Direct Evidence of
             Polaron Formation in Manganites},
  journal = {npj Quantum Mater.},
  year    = {2016},
  volume  = {1},
  pages   = {16026},
  doi     = {10.1038/npjquantmats.2016.26}
}

@article{Seiler_2023,
  author  = {Seiler, Hélène and Zahn, Daniela and Taylor, Victoria C.~A.
             and Bodnarchuk, Maryna I. and Windsor, Yoav W. and Kovalenko, Maksym V.
             and Ernstorfer, Ralph},
  title   = {Direct Observation of Ultrafast Lattice Distortions during
             Exciton--Polaron Formation in Lead Halide Perovskite Nanocrystals},
  journal = {ACS Nano},
  year    = {2023},
  volume  = {17},
  number  = {3},
  pages   = {1979--1988},
  doi     = {10.1021/acsnano.2c06727}
}

@article{Kim_2019,
  author  = {Kim, Tae~Wu and Jun, Sunhong and Ha, Yoonhoo and Yadav, Rajesh~K.
             and Kumar, Abhishek and Yoo, Chung-Yul and Oh, Inhwan and Lim, Hyung-Kyu
             and Shin, Jae~Won and Ryoo, Ryong and Kim, Hyungjun and Kim, Jeongho
             and Baeg, Jin-Ook and Ihee, Hyotcherl},
  title   = {Ultrafast Charge Transfer Coupled with Lattice Phonons in
             Two-Dimensional Covalent Organic Frameworks},
  journal = {Nat. Commun.},
  year    = {2019},
  volume  = {10},
  number  = {1},
  pages   = {1873},
  doi     = {10.1038/s41467-019-09872-w}
}

@article{DeSio_2016,
  author  = {De Sio, Antonietta and Troiani, Filippo and Maiuri, Margherita
             and R{\'e}hault, Julien and Sommer, Ephraim and Lim, James
             and Huelga, Susana~F. and Plenio, Martin~B. and Rozzi, Carlo~Andrea
             and Cerullo, Giulio and Molinari, Elisa and Lienau, Christoph},
  title   = {Watching the Coherent Birth of Polaron Pairs in Conjugated Polymers},
  journal = {Nat. Commun.},
  year    = {2016},
  volume  = {7},
  pages   = {13742},
  doi     = {10.1038/ncomms13742}
}

@article{Wang_2023,
  author  = {Wang, Hui-Min and Liu, Xin-Bao and Hu, Shi-Qi and Chen, Da-Qiang
             and Chen, Qing and Zhang, Cui and Guan, Meng-Xue and Meng, Sheng},
  title   = {Giant Acceleration of Polaron Transport by Ultrafast
             Laser-Induced Coherent Phonons},
  journal = {Sci. Adv.},
  year    = {2023},
  volume  = {9},
  number  = {33},
  pages   = {eadg3833},
  doi     = {10.1126/sciadv.adg3833}
}

@article{Chen_2025,
  author  = {Chen, Wenfan and Wang, Tian and Yu, Chun-Chieh and Jing, Yuancheng
             and Li, Xiaosong and Xiong, Wei},
  title   = {Small Polaron--Induced Ultrafast Ferroelectric Restoration in {$\rm BiFeO_3$}},
  journal = {Phys. Rev. X},
  year    = {2025},
  volume  = {15},
  number  = {2},
  pages   = {021046},
  doi     = {10.1103/PhysRevX.15.021046}
}

@article{lafuente_abinitio_2022,
  title = {Ab initio self-consistent many-body theory of polarons at all couplings},
  author = {Lafuente-Bartolome, Jon and Lian, Chao and Sio, Weng Hong and Gurtubay, Idoia G. and Eiguren, Asier and Giustino, Feliciano},
  journal = {Phys. Rev. B},
  volume = {106},
  issue = {7},
  pages = {075119},
  numpages = {20},
  year = {2022},
  month = {Aug},
  publisher = {American Physical Society},
  doi = {10.1103/PhysRevB.106.075119},
  url = {https://link.aps.org/doi/10.1103/PhysRevB.106.075119}
}

@article{Cinquanta_2019,
  author  = {Cinquanta, Eugenio and Meggiolaro, Daniele and Motti, Silvia G. 
             and Gandini, Marina and Alcocer, Marcelo J. P. and Akkerman, Quinten A.
             and Vozzi, Caterina and Manna, Liberato and De Angelis, Filippo 
             and Petrozza, Annamaria and Stagira, Salvatore},
  title   = {Ultrafast {THz} Probe of Photoinduced Polarons in Lead‐Halide Perovskites},
  journal = {Phys. Rev. Lett.},
  year    = {2019},
  volume  = {122},
  number  = {16},
  pages   = {166601},
  doi     = {10.1103/PhysRevLett.122.166601}
}

@article{Cannelli_2021,
  author  = {Cannelli, Oliviero and Colonna, Nicola and Puppin, Michele
             and Rossi, Thomas C. and Kinschel, Dominik and Leroy, Ludmila M.\ D.
             and Löffler, Janina and Budarz, James M. and March, Anne M. 
             and Doumy, Gilles and others},
  title   = {Quantifying Photoinduced Polaronic Distortions in Inorganic Lead Halide Perovskite Nanocrystals},
  journal = {J. Am. Chem. Soc.},
  year    = {2021},
  volume  = {143},
  number  = {24},
  pages   = {9048--9059},
  doi     = {10.1021/jacs.1c02403}
}

@misc{sup,
  title        = {{See Supplemental Material at}
                  \url{https://link.aps.org/supplemental/10.1103/PhysRevLett.XXX.YYYY}
                  {\,for [brief description of what the supplement contains]}
                 and additional citations to {Refs}.~\cite{PhysRevB.88.085117,PhysRevLett.77.3865,marzari2012maximally,ShengBTE_2014,CarusoPRB23}
                 },
  howpublished = {Supplemental Material},
}

@article{BaroniRMP,
  title = {Phonons and related crystal properties from density-functional perturbation theory},
  author = {Baroni, Stefano and de Gironcoli, Stefano and Dal Corso, Andrea and Giannozzi, Paolo},
  journal = {Rev. Mod. Phys.},
  volume = {73},
  issue = {2},
  pages = {515--562},
  numpages = {0},
  year = {2001},
  month = {Jul},
  publisher = {American Physical Society},
  doi = {10.1103/RevModPhys.73.515},
  url = {https://link.aps.org/doi/10.1103/RevModPhys.73.515}
}

@article{giannozzi2017advanced,
  title={Advanced capabilities for materials modelling with {{\tt Quantum ESPRESSO}} },
  author={Giannozzi, Paolo and Andreussi, Oliviero and Brumme, Thomas and Bunau, Oana and Nardelli, M Buongiorno and Calandra, Matteo and Car, Roberto and Cavazzoni, Carlo and Ceresoli, Davide and Cococcioni, Matteo and others},
  journal={J. Phys. Condens. Matter},
  volume={29},
  number={46},
  pages={465901},
  year={2017},
  url = {https://iopscience.iop.org/article/10.1088/1361-648X/aa8f79/meta},
  publisher={IOP Publishing}
}

@misc{Dai2025,
      title={Comparison between first-principles supercell calculations of polarons and the ab initio polaron equations}, 
      author={Zhenbang Dai and Donghwan Kim and Jon Lafuente-Bartolome and Feliciano Giustino},
      year={2025},
      eprint={2511.01764},
      archivePrefix={arXiv},
      primaryClass={cond-mat.mtrl-sci},
      url={https://arxiv.org/abs/2511.01764}, 
}

@article{Sadigh2015,
  title = {Variational polaron self-interaction-corrected total-energy functional for charge excitations in insulators},
  author = {Sadigh, Babak and Erhart, Paul and \AA{}berg, Daniel},
  journal = {Phys. Rev. B},
  volume = {92},
  issue = {7},
  pages = {075202},
  numpages = {10},
  year = {2015},
  month = {Aug},
  publisher = {American Physical Society},
  doi = {10.1103/PhysRevB.92.075202},
  url = {https://link.aps.org/doi/10.1103/PhysRevB.92.075202}
}

@article{Bretschneider_2018,
  title={Quantifying polaron formation and charge carrier cooling in lead-iodide perovskites},
  author={Bretschneider, Simon A and Ivanov, Ivan and Wang, Hai I and Miyata, Kiyoshi and Zhu, Xiaoyang and Bonn, Mischa},
  journal={Advanced Materials},
  volume={30},
  number={29},
  pages={1707312},
  year={2018},
  publisher={Wiley Online Library}
}

@article{Wu_2021,
  title={Strong self-trapping by deformation potential limits photovoltaic performance in bismuth double perovskite},
  author={Wu, Bo and Ning, Weihua and Xu, Qiang and Manjappa, Manukumara and Feng, Minjun and Ye, Senyun and Fu, Jianhui and Lie, Stener and Yin, Tingting and Wang, Feng and others},
  journal={Science Advances},
  volume={7},
  number={8},
  pages={eabd3160},
  year={2021},
  publisher={American Association for the Advancement of Science}
}
\end{document}